\documentclass[letterpaper]{article} 
\usepackage{aaai23}  
\usepackage{times}  
\usepackage{helvet}  
\usepackage{courier}  
\usepackage[hyphens]{url}  
\usepackage{graphicx} 
\def\UrlFont{\rm}  
\usepackage{natbib}  
\usepackage{caption} 
\usepackage{color}
\usepackage{amsmath}
\usepackage{amsfonts}
\usepackage{subfigure}
\usepackage{algorithm}
\usepackage{algorithmic}

\usepackage{newfloat}
\usepackage{listings}
\DeclareCaptionStyle{ruled}{labelfont=normalfont,labelsep=colon,strut=off} 
\floatstyle{ruled}
\newfloat{listing}{tb}{lst}{}
\floatname{listing}{Listing}
\title{Multi-View MOOC Quality Evaluation via Information-Aware \\ Graph Representation Learning}

\author{
Lu Jiang$^{1}$, Yibin Wang$^{1}$, Jianan Wang$^{1}$\thanks{Corresponding author}, Pengyang Wang$^{2*}$, Minghao Yin$^{1,3*}$\\}
\affiliations{
$^1$School of Computer Science and Information Technology, Northeast Normal University, China\\
$^2$Department of Computer and Information Science, University of Macau, China\\
$^3$Key Laboratory of Applied Statistics of MOE, Northeast Normal University, Changchun, China\\
\{jiangl761, wangyb856,wangjn\}@nenu.edu.cn, pywang@um.edu.mo,     ymh@nenu.edu.cn
}

\usepackage{bibentry}

\begin{document}

\maketitle

\begin{abstract}
In this paper, we study the problem of MOOC quality evaluation which is essential for improving the course materials, promoting students' learning efficiency, and benefiting user services. 
While achieving promising performances, current works still suffer from the complicated interactions and relationships of entities in MOOC platforms. 
To tackle the challenges, we formulate the problem as a course representation learning task-based and develop an \textbf{I}nformation-\textbf{a}ware \textbf{G}raph \textbf{R}epresentation \textbf{L}earning(\textbf{IaGRL}) for multi-view MOOC quality evaluation. 
Specifically, We first build a MOOC Heterogeneous Network (HIN) to represent the interactions and relationships among entities in MOOC platforms. 
And then we decompose the MOOC HIN into multiple single-relation graphs based on meta-paths to depict the multi-view semantics of courses. 
The course representation learning can be further converted to a multi-view graph representation task. 
Different from traditional graph representation learning, the learned course representations are expected to match the following three types of validity: 
(1) the agreement on expressiveness between the raw course portfolio and the learned course representations; 
(2) the consistency between the representations in each view and the unified representations; 
(3) the alignment between the course and MOOC platform representations. 
Therefore, we propose to exploit mutual information for preserving the validity of course representations. 
We conduct extensive experiments over real-world MOOC datasets to demonstrate the effectiveness of our proposed method.
\end{abstract}

\section{Introduction}

Massive open online course (MOOC) has been prevalent for online tutoring and self-studying in recent decades by providing numerous course materials, intermediate feedback, and interactions between student and instructors. 
Among which, MOOC course quality evaluation is one of the vital tasks in MOOC platform management for helping improve the course materials, promote students' learning efficiency~\cite{jiang2021eduhawkes}, and benefit user services ({\it e.g.}, course recommendation~\cite{DBLP:journals/tois/WangZWZZCX22,article}, cognitive diagnosis~\cite{jiang2022augmenting}, etc). 

Current studies in MOOC quality evaluation lie in two aspects:
(1) manual evaluation~\cite{DBLP:journals/iahe/WangLLMY21}, which evaluates the course quality by domain experts based on a pre-defined rubric;
and 
(2) automated evaluation~\cite{DBLP:journals/uais/Perez-MartinRM21,DBLP:journals/aai/BetanzosCB17}, which judges the course quality automatically based on historical records in the platform. 
While achieving promising results, current works still exhibit limitations: 
First of all, manual-based methods are time-consuming and labor-intensive.
And, in most of cases, the experts do not have the complete domain knowledge to evaluate every course on the MOOC platform.
Second, most of the automated methods utilize students' reviews as the criteria for evaluating course quality.
However, students' evaluation of the quality of MOOCs is biased and subjective, and cannot yield unified evaluations.
In fact, course quality evaluation in the MOOC platform is a complicated process involving by multiple parties, which can be examined from different views. 
Therefore, integrating semantics and opinions from different views to inform unified representations of the course becomes the key to reasonable MOOC quality evaluation.

However, two unique challenges arise in achieving this goal.
First, how to design an appropriate data structure for capturing complex interactions among different types of entities in the MOOCs platform?
Second, how should we guarantee the validity of the multi-view representations?
Next, we will outline how we tackle these challenges.

First, in a MOOCs platform, we observe that in addition to the student and course, there exist multiple types of entities and multiple types of relationships between pairs of different entities. 
The complex MOOC data structures are always represented in heterogeneous information networks (MOOC HIN)~\cite{DBLP:journals/tkde/ShiLZSY17}. 
Considering the participation of multiple entities on the MOOC platform and the impact of interactions between entities and courses on MOOC quality evaluation.
It is difficult to obtain a comprehensive evaluation of the MOOC quality if we merely depend on a single view.
Only utilizing a single type of interaction may overlook important relationships between courses and other entities.
For example, "student click course" and "teacher upload course" have dissimilar semantics even though they are included in the same course.
These heterogeneous relationships provide rich information from multi-view.
Thus, it requires incorporating these heterogeneous relationships into the representation learning of the course entities.
To address the above issues, we use meta-paths~\cite{DBLP:journals/tois/SunYXMZ17} as the guidance to capture multi-view representations of courses in MOOC heterogeneous information network.

Second, although multi-view node embedding can be obtained by performing representation learning on MOOC HIN, how guaranteeing the validity of the learned course representations remains a challenge. 
Specifically, the validity of course representations lies in three aspects: 
(1) the course representations should preserve the same semantics as the raw course portfolio; 
(2) the representations in each view should be consistent with the unified representations of the course; 
(3) the course representations should be aligned with the overall representations of the MOOC platform. 
The three types of validity indicate strong correlations between the three pairs of representations. 
Therefore, to ensure validity, we aim to maximize the three correlations between the pair of course representations and the raw course portfolio, the pair of unified course representations and each view, and the pair of course representations and platform representations.
In this paper, we exploit mutual information (MI), a powerful correlation measure, to quantify the correlations in each pair of representations.

In summary, we propose an \textbf{I}nformation-\textbf{a}ware \textbf{G}raph \textbf{R}epresentation \textbf{L}earning(\textbf{IaGRL}) for multi-view MOOC quality evaluation.
The main contributions are as follows:
\begin{itemize}
    \item We formulate the problem of MOOC quality evaluation as a multi-view graph representation learning task.
    \item We construct MOOC HIN and propose to exploit meta-paths to extract the semantics of MOOC relationships in different views.
    \item We identify three types of validity of course representations, and provide an information-aware course representation learning framework.
    \item We conduct extensive experiments over real-world MOOC datasets to validate the effectiveness of our proposed method.
\end{itemize}

\section{Definitions and Problem Statement}

We introduce the key definitions and the problem statement.
Then we present the overview of the proposed method. 
Some important notations are summarized in Table~\ref{table_notation}.

\begin{table}[htbp]
\scriptsize
	\tabcolsep 0.08in
\caption{Symbol and Definitions}
\begin{center}
\begin{tabular}{c|c}
\hline
\textbf{Symbol} & \textbf{Definition}\\
\hline
$\mathcal{G}$ & A given MOOC heterogeneous graph\\

$\mathcal{V,E}$ & Set of nodes, edges\\

$v,e$ & MOOC HIN node, edge\\

$\mathcal{MP}$ & A set of meta-paths \\

$\mathbf{X}$ & The features matrix of courses \\

$\mathbf{A}$ & The adjacency matrix base different meta-paths\\

$\Tilde{\textbf{D}}$ & The diagonal matrix base different meta-paths\\

$\Tilde{\textbf{h}}$ & The multi-view course representation\\

$\textbf{W}$ & The weights of GCN layer\\

$\textbf{h}$ & The unified course representation\\

$\alpha^{MP_i}$ & The importance of each meta-path\\

$\textbf{M}$ & The platform representation\\

$\mathcal {D}$ & Mutual information based discriminator\\

$\lambda_q, \lambda_j, \lambda_s, \lambda_y$ & The weight of different losses\\

\hline
\end{tabular}
\label{table_notation}
\end{center}
\end{table}

\subsection{Definitions and Problem Statement}

\newtheorem{definition}{Definition}
\begin{definition}
{\bf MOOC Heterogeneous Information Network(HIN)}
A MOOC HIN is defined as $\mathcal{G} = (\mathcal{V}, \mathcal{E})$ with a node type mapping function $\phi:\mathcal{V} \rightarrow \mathcal{N}$ and an edge type mapping function $\psi:\mathcal{E} \rightarrow \mathcal{R}$.
Specifically, in MOOC HIN, there are four types of nodes: students(denoted as U), teachers(denoted as T), courses(denoted as C), and subjects(denoted as S).
And there are three types of links: "click" which is to demonstrate the relation between students and courses, "upload" which is to demonstrate the relation between teachers and courses, and "include" which is to demonstrate the relation between subjects and courses.
The MOOC HIN is defined as the following groups of triplet facts:(1) $<$student, $"$click$"$, course$>$, (2) $<$teacher, $"$upload$"$, course$>$, and (3) $<$subject, $"$include$"$, course$>$. 
\end{definition}

\begin{definition}
{\bf Meta-path based on MOOC HIN}
A meta-path $MP$ based on MOOC HIN is defined as a path in the form of $\mathcal{V}_1 \stackrel{\mathcal{E}_1}\to \mathcal{V}_2 \stackrel{\mathcal{E}_2}\to \cdots \stackrel{\mathcal{E}_l}\to \mathcal{V}_{l+1}$(abbreviated as $\mathcal{V}_1, \mathcal{V}_2,\cdots,\mathcal{V}_{l+1}$), which describes a composite relation $\mathcal{E} = \mathcal{E}_1 \circ \mathcal{E}_2 \circ \cdots \circ \mathcal{E}_l$ between object $\mathcal{V}_1, \mathcal{V}_2, \cdots, \mathcal{V}_{l+1}$, where $\circ$ denotes the composition operator on relations. 
In the MOOC HIN, two courses can be connected via multiple paths, {\it e.g.}, $C \stackrel{upload}\longrightarrow T \stackrel{upload^{-1}}\longrightarrow C$, 
$C \stackrel{click}\longrightarrow U \stackrel{click^{-1}}\longrightarrow C$,
$C \stackrel{include}\longrightarrow S \stackrel{include^{-1}}\longrightarrow C$. 
A set of meta-paths are defined as $\mathcal{MP}= \{ MP_1, MP_2, \cdots MP_{||MP||} \}$.

\end{definition}

\begin{definition}
{\bf Problem Statement}
In this paper, we study the problem of MOOC quality evaluation.
We formulate the problem as an information-aware graph representation learning task.
Formally, we aim to find a mapping function $f : \mathcal {G} \rightarrow \textbf{h}$ that takes the MOOC HIN $\mathcal {G}$ as input, and outputs information-aware representations $\textbf{h}=\{ \textbf{h}_1,\textbf{h}_2,...,\textbf{h}_N \}$, for evaluating the course quality on a MOOC platform.

\end{definition}

\begin{figure*}[!t]
	\centering
	\includegraphics[width=1.0\linewidth]{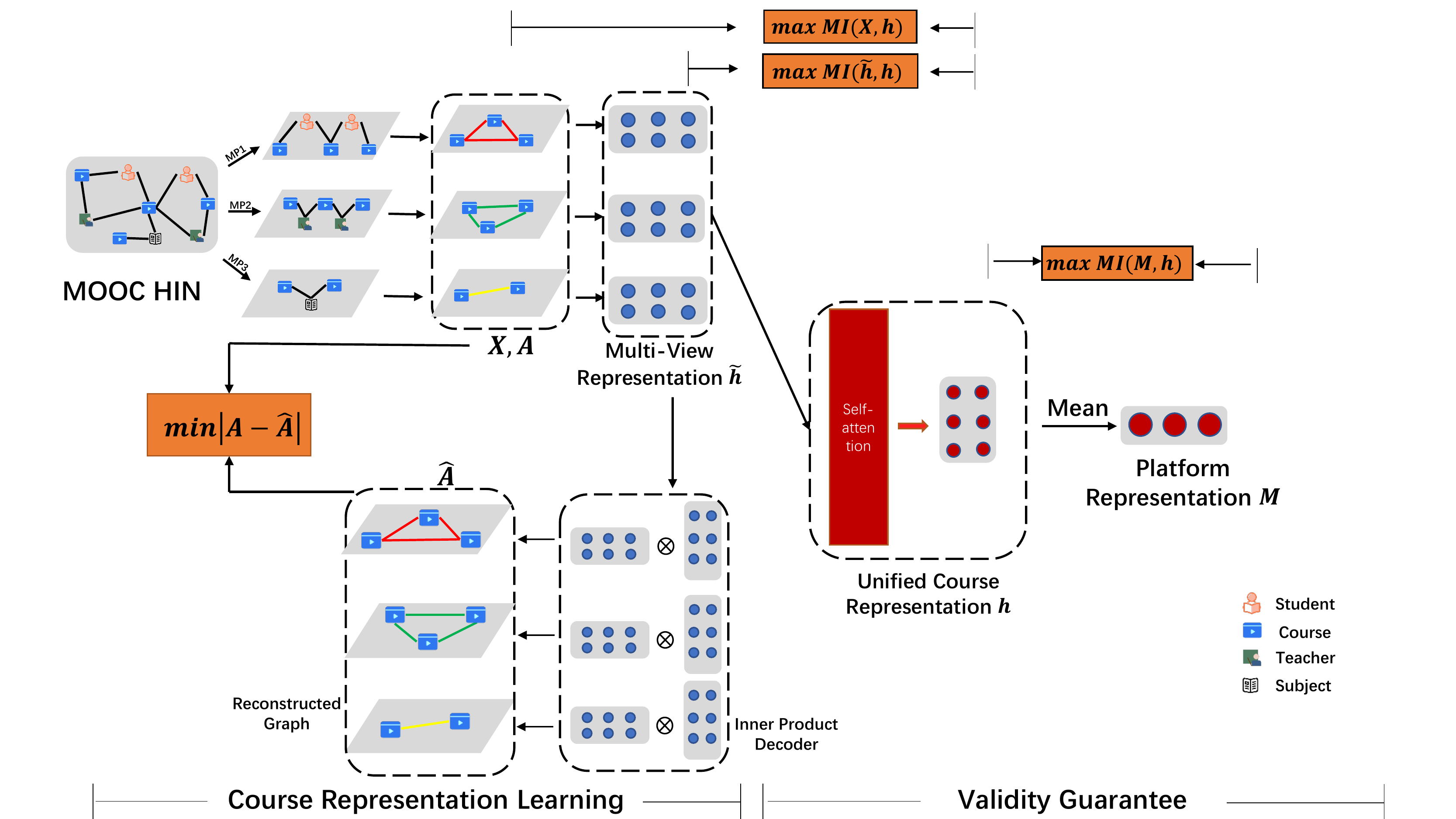}
	\captionsetup{justification=centering}
	\caption{Framework Overview.}
	\label{fig:framework overview}
\end{figure*}

\section{Method}
In this section, we first present an overview of our proposed framework. 
Then, we introduce the multi-view course representation learning with validity guarantee in detail.

\subsection{Framework Overview}
Figure~\ref{fig:framework overview} shows an overview of the proposed two-stage framework: 
(1) Stage 1: multi-view course representation of learning; and 
(2) Stage 2: course representation validity guarantee. 
Specifically, in Stage 1, we extracted our three meta-paths from the MOOC HIN and constructed the course adjacency matrix under three different views through the meta-path. 
Then, we proposed to learn the representation of course under different views by using the graph convolutional network(GCN)~\cite{DBLP:journals/corr/KipfW16} through the encoder-decoder paradigm. 
By minimizing the reconstruction loss that follows the convention of contrastive learning styles. 
In Stage 2, we exploit MI maximization to ensure the three types of course representation validity, with (i) the agreement on expressiveness between the raw course portfolio and the learned course representations; 
(ii) the consistency between the representations in each view and the unified representations; and 
(iii) the alignment between the course and MOOC platform representations. 

\subsection{Course Representation Learning}

\subsubsection{Multi-View Representations.}

Given the MOOC HIN $\mathcal{G =(V, E)}$ with a set of meta-paths $\mathcal{MP}= \{ MP_1, MP_2, \cdots MP_{||MP||} \}$ and the corresponding adjacency matrix $\mathbf{A}= \{A_1, A_2, \cdots A_{|MP|}\}$, and $|MP|$ denotes the number of meta-paths.
Let $\mathbf{X}= \{X_1, X_2, \cdots X_n\}$ denoted as the course attribute matrix of the MOOC HIN.
In this paper, the course attributes are represented by a $d$-dimensional vector that describes the contents of the course, including the headline, abstract, etc.
The course attribute matrix is generated through the Doc2Vec~\cite{DBLP:conf/icml/LeM14} model.

For better generality, we learn multi-view representation with GCN in an unsupervised fashion.
We use generalized advantage estimation(GAE)~\cite{DBLP:journals/corr/SchulmanMLJA15} to learn representation in an encode-decode paradigm.
Specifically, the encoder is a GCN with the following layer-wise propagation rule, the multi-view course representation can be denoted as:

\begin{equation}
  \Tilde{\textbf{h}} =(\Tilde{\textbf{D}}^{-\frac{1}{2}}{\Tilde{\textbf{A}}}\Tilde{\textbf{D}}^{-\frac{1}{2}})\textbf{X}\textbf{W},
  \label{eq1}
\end{equation}
where $\Tilde{\textbf{A}} = \textbf{A} + \textbf{I}$ is the adjacency matrix  corresponds to a single meta-path with self-connections and $\textbf{I}$ is the identity matrix.   
$\Tilde{\textbf{D}}_{ii} = \sum_{i} \Tilde{\textbf{A}}_{ij}$ is the diagonal matrix, and $\textbf{W}$ is the weight. 
The decoder is an inner product of the learned representation to recover the adjacency matrix:

\begin{equation}
    \hat{\textbf{A}} = sigmoid(\Tilde{\textbf{h}} {\Tilde{\textbf{h}}^T})
     \label{eq2}
\end{equation}

The objective is to minimize the reconstruction loss between the original adjacency matrix $\Tilde{\textbf{A}}$ and reconstructed adjacency matrix $\hat{\textbf{A}}$.
We follow the implementation of VGAE~\cite{DBLP:journals/corr/KipfW16a} to do the sampling and loss optimization: we take connected neighbors as positive nodes, and disconnected nodes as negative nodes, and sample a few of them to construct the data samples. 
We expect positive samples to be connected, and negative samples to be disconnected after reconstruction, thus the reconstruction is converted to a classification task, which can be optimized using cross-entropy loss.


\begin{equation}
    \mathcal{L}_q = -log\hat{\textbf{A}}_{pos}-log(1-\hat{\textbf{A}}_{neg}),
    \label{eq3}
\end{equation}
where $\hat{\textbf{A}}_{pos}$ and $\hat{\textbf{A}}_{neg}$ are derived from the positive course nodes pairs and the negative course node pairs respectively, based on Equation~\ref{eq3}.

\subsubsection{Unified Course Representation.}

Going through the GAE, we learn the representations for each meta-path.
However, different meta-paths should not be considered equally.
To address this problem, we adopt the idea of heterogeneous deep graph infomax(HDGI)~\cite{DBLP:journals/corr/abs-1911-08538}, and utilize the self-attention mechanism to fuse the embedding of courses learned under the guide of different meta-paths and generate the unified course embedding.
Specifically, we learn the self-attention weights for different meta-paths as follows:

\begin{equation}
    \textbf{h}=  \sum_i^{|MP|} att(\tilde{\textbf{h}}_i),
    \label{eq4}
\end{equation}
where att($\cdot$) indicates the self-attention function,
and $\textbf{h}$ indicates the unified course representation, which has integrated the self-attention weights of different meta-paths. 
In this paper, we mainly focus on the course from a multi-view.
In order to make representations from different meta-paths comparable, we transform each course's representation from a different view with a linear transformation.
The parameters are shared weight matrix $\textbf{W}'$ and shared bias vector $\vec{b}$.
Based on the distinguishing ability of views, we introduce the shared attention vector $\vec{q}$ of different views to calculate the importance of each view.
The importance of the meta-paths can be calculated as follows:

\begin{equation}
a^{MP_i} = \frac{1}{|MP|}\sum^{|MP|}_{j=1}
tanh(\vec{q}^T\cdot[\textbf{W}'\cdot\tilde{\textbf{h}}^{{MP}_i}_j+\vec{b}])
\label{eq5}
\end{equation}

Then, we use the softmax function to normalize the importance of meta-paths, the normalized weight of each meta-path can be calculated as follows:

\begin{equation}
\alpha^{MP_i}= \frac{exp(a^{MP_i})}{\sum_{j=1}^{|MP|}exp(a^{MP_j})}
\label{eq6}
\end{equation}

The self-attention unified course representation $\textbf{h}$ can be represented as follows:

\begin{equation}
\textbf{h}= \sum_{i=1}^{|MP|}\alpha^{MP_i}\tilde{\textbf{h}}_i
\label{eq7}
\end{equation}

\subsection{Validity Guarantee}

\subsubsection{Raw Portfolio-Representation Agreement.} 
The learned course representations are expected to achieve agreement with the raw course portfolio in describing courses.
We refer to this as Raw Portfolio-Representation Agreement.
We use mutual information (MI) to quantify the agreement between the representation $\textbf{h}$ learned by the course node in the unified view and the representation $\textbf{X}$ of the raw features of the MOOC. 
Following the idea of DIM and DGI~\cite{DBLP:conf/iclr/VelickovicFHLBH19}, a Jensen Shannon MI estimator is defined to estimate and maximize the MI between $\textbf{X}$ and $\textbf{h}$:

\begin{equation}
\begin{aligned}
    MI(\textbf{X}, \textbf{h}) := &\mathbb{E}_{\textbf{X}}[-sp(-\mathcal{D}(\textbf{X}, \textbf{h}^{(MP)}))]+\\
    &\mathbb{E}_{\bar{\textbf{X}}}[sp(\mathcal{D}(\textbf{X}, \textbf{h}^{(MP)}))],\\
\end{aligned}
\label{eq8}
\end{equation}
where sp is the softplus function that $sp(c)\times log(1+e^c)$, $\textbf{X}$ is the positive sample set and $\bar{\textbf{X}}$ is negative sample set.
We will present how we generate positive and negative samples later.
Since the noise-contrastive type objective with a standard binary cross-entropy (BCE) can effectively maximize mutual information, we define the loss function as:

\begin{equation}
\begin{aligned}
    \mathcal{L}_j = -\frac{1}{|MP|}\sum_{j=1}^{|MP|}\sum_{i}&\mathbb{E}_{\textbf{X}}[log\mathcal{D}_j(\textbf{X}^{({MP}_j)}_i,{\textbf{h}}_i)]-\\
    &\mathbb{E}_{\bar{\textbf{X}}}[log(1-\mathcal{D}_j(\Tilde{\textbf{X}}_i^{({MP}_j)},{\textbf{h}}_i))],
    \label{eq9}
\end{aligned}
\end{equation}
where $\mathcal{D}_j$ denotes a discriminator to justify the given pairs as positive or negative.
For the $k$-th view, we regard the positive sample as the pair of $(X^{({MP}_j)}_i, {\textbf{h}}_i)$ and the negative samples as the pairs of $(\Tilde{X}^{({MP}_j)}_i, {\textbf{h}}_i)$.

\subsubsection{Multi-View Consistency.} 
Although we deconstruct heterogeneous graphs into different views, the course representations in each view are expected to be consistent with the unified course representations in semantics, which is defined as multi-view consistency. 
We propose to exploit MI to measure the multi-view consistency.
First, we get the  multi-view representation $\tilde{\textbf{h}}$ and unified course representation $\textbf{h}$. 
Specifically, $\textbf{h}$ is obtained from self-attention on the one hand and $\tilde{\textbf{h}}$ is obtained from the GCN encoder.
Then, we use neural network estimation MI $({\textbf{h}}; \tilde{\textbf{h}})$ to maximize the mutual information between the unified course representation $\textbf{h}$ and multi-view representation $\tilde{\textbf{h}}$.
We have a similar noise contrastive loss function:

\begin{equation}
\begin{aligned}
    \mathcal{L}_s = -\frac{1}{|MP|}\sum_{j=1}^{|MP|}\sum_{i}&\mathbb{E}_{\textbf{X}}[log\mathcal{D}_s({\textbf{h}}_i,\tilde{\textbf{h}}_i^{(MP_j)})]-\\
    &\mathbb{E}_{\bar{\textbf{X}}}[log(1-\mathcal{D}_s({\textbf{h}}_i,\Acute{\textbf{h}}_i^{({MP}_j)}))],
    \label{eq10}
\end{aligned}
\end{equation}
where $\mathcal{D}_s$ denotes a discriminator for discriminating positive consistent pairs, ${\textbf{h}}_i$ is unified course representation.$\Acute{\textbf{h}_i}$ is the result of $\tilde{\textbf{h}}$ after random shuffling.
For the $k$-th view, we design the positive samples as the pairs of $({\textbf{h}}_i,\tilde{\textbf{h}}_i^{({MP}_j)})$, and the negative samples as the pairs of $({\textbf{h}}_i,\Acute{\textbf{h}}_i^{({MP}_j)})$.
The objective is to minimize $\mathcal{L}_s$, which is equivalent to maximize $MI({\textbf{h}},\tilde{\textbf{h}})$.

\subsubsection{Course-Platform Alignment.}
While there is no doubt that the courses are different from each other, the course representations are required to align with the MOOC platform representations within the same semantic scope.
To accomplish the course-platform alignment, we first obtain the platform representation by considering it as the graph-level representation of MOOC HIN for courses. 
Along this line, we take the platform summary vector $\textbf{M}$ by averaging over all course representations:

\begin{equation}
    \textbf{M} = \sigma(\frac{1}{N}\sum^N_{i=1}{\textbf{h}}_i), 
    \label{eq11}
\end{equation}
where $\sigma$ is the sigmoid function and $N$ is the number of course nodes.
We continue to leverage MI to capture the course-platform alignment, by maximizing the MI between 
In order to maximize course-platform alignment, we introduce MI and then based on the relationship between the Jensen-Shanno degree and mutual information. 
We can maximize the mutual information between platform representation and unified course representation using the binary cross-entropy loss of the discriminator as follows:

\begin{equation}
\begin{aligned}
    \mathcal{L}_y = -\sum_{i}&\mathbb{E}_{\textbf{X}}[log\mathcal{D}_y({\textbf{h}}_i, \textbf{M})]-\\
    &\mathbb{E}_{\bar{\textbf{X}}}[log(1-\mathcal{D}_y(\bar{\textbf{h}}_i, \textbf{M}))],
    \label{eq12}
\end{aligned}
\end{equation}
where $\mathcal{D}_y$ denotes a discriminator to provide probability scores for sampled course-platform pairs, $\textbf{h}$ is unified course representation.$\bar{\textbf{h}}$ is the result of ${\textbf{h}}$ after random shuffling. 
We design the positive samples as the pairs of $({\textbf{h}}_i, \textbf{M})$, and the negative samples as the pairs of $(\bar{\textbf{h}}_i, \textbf{M})$.
The objective is to minimize  $\mathcal{L}_y$, which is equivalent to maximize $MI(\textbf{h}, \textbf{M})$.
\subsection{Optimization}
The loss of the model includes: 
(i) the contrastive learning loss for graph reconstruction $\lambda_q$ (Equation~\ref{eq3}); 
(ii) the raw portfolio-representation agreement learning loss $\lambda_j$ (Equation~\ref{eq9}); 
(iii) the multi-view consistency learning loss  $\lambda_s$ (Equation~\ref{eq10}); 
and (iv) the course-platform alignment learning loss $\lambda_y$ (Equation~\ref{eq11}). 
The objective is to minimize the overall loss $\mathcal{L}$ as follows:
\begin{equation}
\begin{aligned}
    \mathcal{L} =
    \lambda_q\mathcal{L}_q+\lambda_j\mathcal{L}_j+\lambda_s\mathcal{L}_s+\lambda_y\mathcal{L}_y
    \label{eq15}
\end{aligned}
\end{equation}
where $\lambda_q$, $\lambda_j$, $\lambda_s$ and $\lambda_y$ are the weights for $\mathcal{L}_    q$, $\mathcal{L}_j$, $\mathcal{L}_s$ and $\mathcal{L}_y$, respectively.
The above loss can be optimized through gradient descent, and the representations of nodes can be learned when the optimization is completed.

\section{Experiment}

In the experiment, we aim to answer the following three research questions:
\begin{itemize}

\item {\bf Q1.} How is the performance of our proposed \textbf{IaGRL} in the MOOC quality evaluation task?

\item {\bf Q2.} How do the meta-paths affect the course quality evaluation performance?

\item {\bf Q3.} How do the different learning losses affect the course quality evaluation performance? 

\end{itemize}

Then, we will introduce statistical information about real-world MOOC data, and experiment settings and compare \textbf{IaGRL} with several baselines on this data.

\subsection{Data Description}

We evaluate the performance over real-world MOOC data.
The data constitute a MOOC heterogeneous information network containing 4 types of entities and 3 types of relations.
The course scores range from 0 to 5.
In the data preprocessing step, we filtered out users that have fewer than 3 links.
After data preprocessing, we split the datasets into two non-overlapping sets: 20\% of the datasets as the testing set and the rest 80\% as the training set. 
Table~\ref{table_data} shows the detailed statistics of the dataset.

\begin{table}[htbp]
\caption{Statistics and descriptions of dataset}
\begin{center}
\begin{tabular}{c|c|c}
\hline
Properties & Descriptions & Statistics\\
\hline
Student & Users who studied course&4931  \\
Course & Learning materials &10919  \\
Teacher & Users who uploaded course &1213 \\
Subject & An area of knowledge &35   \\
\hline
Time & Time period & 2015-2018\\ 
\hline
\end{tabular}
\label{table_data}
\end{center}
\end{table}

\subsection{Baselines and Evaluation Metrics}

We compare the performances of our method with the following baselines:

\noindent{\bf (1) MLP.} The multilayer perceptron(MLP) is a feedforward supervised artificial neural network structure. The MLP can contain multiple hidden layers to realize the classification modeling.

\noindent{\bf (2) DeepWalk.} The DeepWalk model is a recently proposed network embedding method that extends the word2vec model~\cite{DBLP:conf/nips/MikolovSCCD13} by truncated random walks~\cite{DBLP:conf/kdd/PerozziAS14}.

\noindent{\bf (3) DeepWalk+F.} The DeepWalk+F model not only adds the neighbor features after the random walk but also adds the attribute features of practice.

\noindent{\bf (4) GCN.} The graph convolutional network(GCN) performs information aggregation based on the Laplacian or adjacent matrix of the complete graph~\cite{DBLP:journals/corr/KipfW16}.

\noindent{\bf (5) GAE.} The graph autoencoder(GAE) learns node representation, we use an embedding layer to encode the node with GCN~\cite{DBLP:journals/corr/SchulmanMLJA15}.

\noindent{\bf (6) GraphSAGE.} The GraphSAGE proposes neighborhood batch sampling to enable scalable training with max, min, and LSTM aggregation functions~\cite{DBLP:conf/nips/HamiltonYL17}.

\noindent{\bf (7) GAT.} The graph attention network(GAT) introduces a multi-head attention mechanism into the aggregation function, which learns the importance of the neighborhood of each node for information aggregation~\cite{DBLP:journals/corr/abs-1710-10903}.

\noindent{\bf (8) GATv2.} GATv2 solves the simple graph problem of GAT using a static attention mechanism, a dynamic graph attention variant that is more expressive than GAT~\cite{DBLP:journals/corr/abs-2105-14491}.

The last five are graph representation learning methods.
To evaluate the performance of the models for MOOC quality, we adopt two widely used evaluation metrics for multi-classification performance, {\it e.g.}, Accuracy and Macro-F1.
Specifically, Accuracy measures the evaluation accuracy that the user scores successfully rated.
And macro f1 help to consider performance comprehensively in case of imbalanced data. 
For both metrics, the larger the value, the better the performance.

\subsection{Parameter Setting}

For Deepwalk and Deepwalk+F, we set the number of walks = 80, the size of representation = 128, the walk length = 20, and the window size = 10.
For GCN, GAE and GraphSAGE, we set the number of layers = 2, the input feature size=128, the output feature size = 128, and the learning rate = 0.001.
For GAT and GATv2, we set the number of layers = 2, the layer heads = [2,1], the input feature size = 128, the output feature size = 128, and the learning rate = 0.003. For my model we set the learning rate = 0.001, the l2 = 0.001,  the dropout = 0.1, the input feature = 128, and the out feature = 128.
The device we used was two RTX 6000 with 24GiB memory and CUDA=11.2.

\subsection{Overall Performance(\bf Q1)}

\begin{figure}[htbp]
\centering
\subfigure[Accuracy.]{
\begin{minipage}[t]{0.498\linewidth}
\centering
\includegraphics[width=1\linewidth]{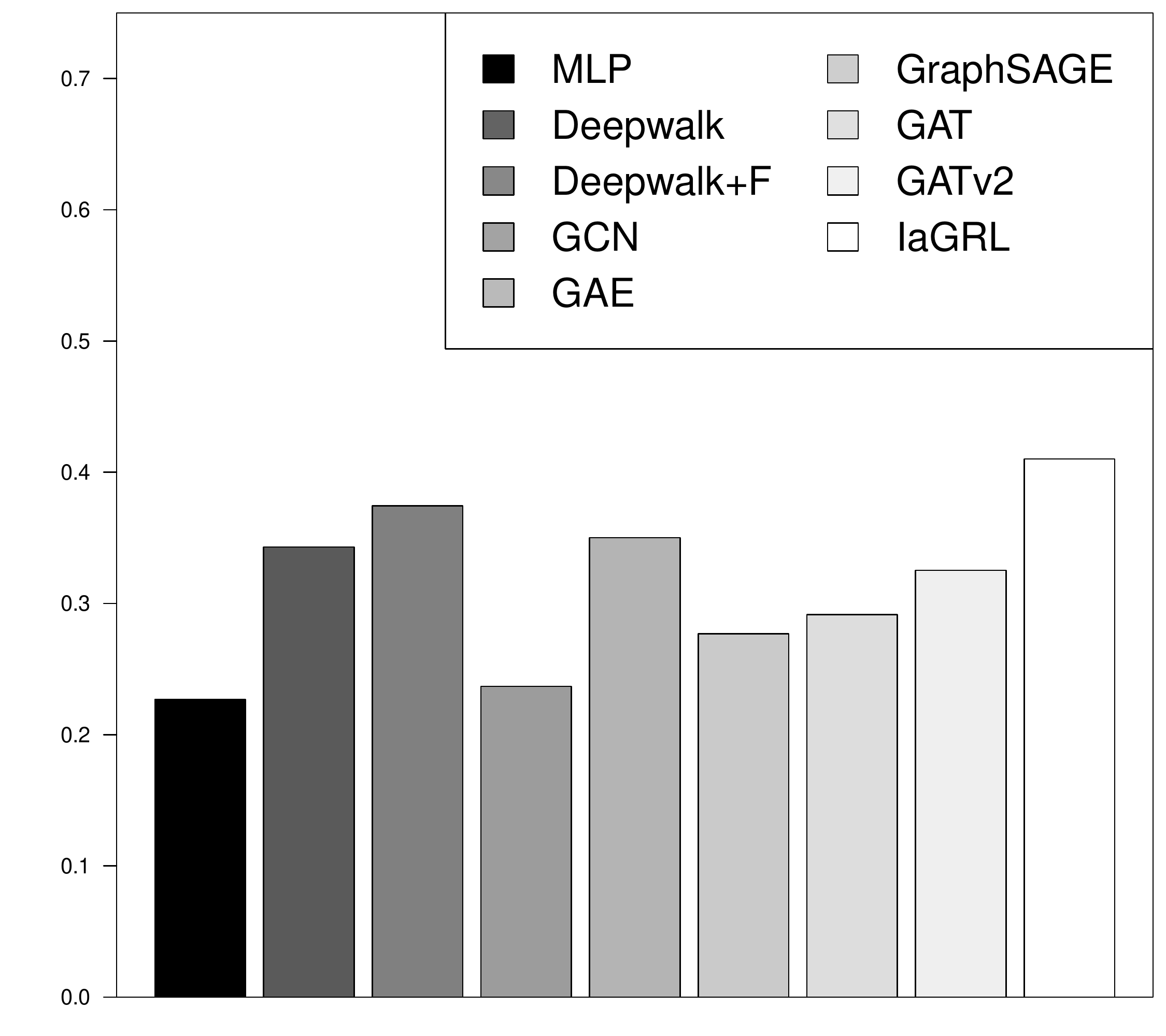}
\label{acc}
\end{minipage}%
}%
\subfigure[Macro-F1.]{
\begin{minipage}[t]{0.498\linewidth}
\centering
\includegraphics[width=1\linewidth]{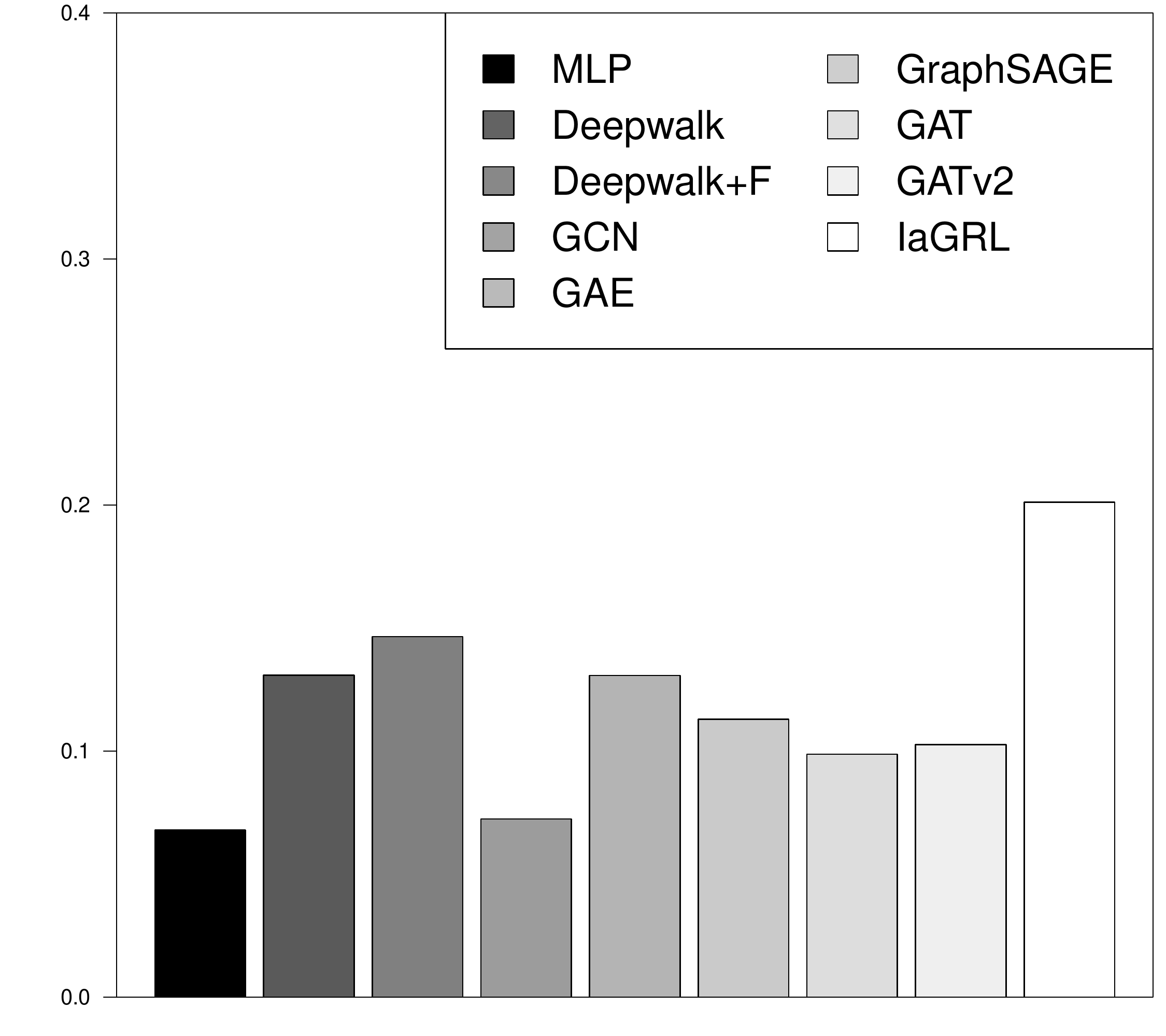}
\label{f1}
\end{minipage}%
}%

\caption{Overall comparisons of evaluation Accuracy and Macro-F1 of our methods with other baselines.}
\label{overallperformance}
\end{figure}

In this section, we compare the overall performance of all models on the real-world dataset.
In general, Figure~\ref{overallperformance} shows our model outperforms other baseline methods for both Accuracy and Macro-F1 metric.
Compared to MLP, which is the representative of node attribute, random walk-based methods (DeepWalk, Deepwalk+F), graph convolution network-based methods (GCN, GAE, GraphSAGE, GAT, GATv2), and information-based methods (our proposed method) perform better in modeling MOOC heterogeneous network.  
Compared to DeepWalk and DeepWalk+F, which is the random walk-based methods, our proposed framework additionally considers heterogeneous graph embedding and information-aware of the learned representation.
Compared to graph convolution network-based methods, the information guarantee provided by our proposed method further elevates the reasonability of the learned representations.
In summary, the results validate that incorporating multi-view and information-aware can improve the quality of representation learning. 

\subsection{Influence of Meta-paths(\bf Q2)}

In this part of the experiments, we analyze how meta-paths affect the performance of methods.
We consider both single meta-path and their combinations in our method.
Specifically, we select three types of meta-paths to represent the relatedness between pair of courses, including MP1: $C \rightarrow T \stackrel{-1}\rightarrow C$, MP2: $C \rightarrow U \stackrel{-1}\rightarrow C$ and MP3: $C \rightarrow S \stackrel{-1}\rightarrow C$.
To analyze the impact of meta-paths, we study the performance in three aspects:(1)with single view attention weights; (2)with single course-related meta-path and their combinations on our method; and (3)with meta-path on baselines.

\subsubsection{Compared with attention weights.}
We calculate the importance of the attention weights for each meta-path in our method, the results are shown in table~\ref{table_attention}.
From the table, we can find the most important meta-path is MP$_1$, from the teacher view, follow by MP$_3$(from the subject view) and MP$_2$(from the student view), respectively.
It is easy to understand that when evaluating a course quality from multiple views, the teacher's influence on course quality is more important.
One interesting observation is that the student's view has the least impact on course quality, even less than the subject view, which is an objective perspective.
A possible explanation is that our data came from a MOOC platform based on primary education.
On the one hand, the student's cognitive level is in the primary state, on the other hand, the students' behavior of clicking courses is guided by the teacher, which is less subjective.

\begin{table}[htbp]
\caption{The importance of each meta-path.}
\begin{center}
\begin{tabular}{c|ccc}
\hline
 Weights & MP$_1$ & MP$_2$ & MP$_3$  \\
\hline
$\alpha^{MP_i}$ & \textbf{0.4655} & 0.2242 & 0.3103  \\
\hline
\end{tabular}
\label{table_attention}
\end{center}
\end{table}

\subsubsection{Compared with the different meta-paths combination.} 
We compare our method with both single meta-path and their combinations. 
The results are shown in Table~\ref{table_combineview}, we can find that every single meta-path exhibits different performance, where the performance ranking is MP1$>$MP3$>$MP2, and the combinations of single meta-paths follow the same tendency. 
This illustrates that different meta-paths indicate different relations and the combination including more meta-paths will exhibit better performance, and the best performance is achieved by combining all three meta-paths.

\begin{table}[htbp]
\caption{The results from different combinations of meta-paths on our method.}
\begin{center}
\begin{tabular}{c|cc}
\hline
Meta-path & Accuracy & Macro-F1 \\
\hline
MP$_1$ & 0.3596 & 0.1529  \\
MP$_2$ & 0.3407 & 0.1397  \\
MP$_3$ & 0.3543 & 0.1707 \\
MP$_1$ \& MP$_2$ & 0.3697 & 0.1536 \\
MP$_1$ \& MP$_3$  & 0.3864 & 0.1783  \\
MP$_2$ \& MP$_3$  & 0.3656 & 0.1742 \\
MP$_1$ \& MP$_2$ \& MP$_3$  & \textbf{0.4101}  & \textbf{0.2011} \\
\hline
\end{tabular}
\label{table_combineview}
\end{center}
\end{table}

\subsubsection{Compared with meta-paths on baselines.}
And in order to further verify the effect of meta-paths, we study the meta-path on the baselines based on graph convolutional network methods.
The results are shown in Table~\ref{table_multiview}, from Table~\ref{table_multiview}, we can find that compared with the original algorithms, including meta-path combinations will show better performance.
Especially in the GCN, and GraphSAGE methods, the growth of performance is quite obvious.

\begin{table}[htbp]
\caption{The results from different combinations of meta-paths on baselines.}
\begin{center}
\begin{tabular}{c|cc}
\hline
Methods& Accuracy & Macro-F1\\
\hline
GCN & 0.2368 & 0.0724 \\
GCN \& MP$_{i=1,2,3}$ & \textbf{0.3466} & \textbf{0.0857} \\
\hline
GAE & 0.3501 & 0.1307\\
GAE \& MP$_{i=1,2,3}$ & \textbf{0.3555} & \textbf{0.1686}\\
\hline
GraphSAGE & 0.2770 & \textbf{0.1130} \\
GraphSAGE \& MP$_{i=1,2,3}$ & \textbf{0.3620} & 0.1092\\
\hline
GAT & 0.2914 & 0.0987 \\
GAT \& MP$_{i=1,2,3}$ & \textbf{0.3187} & \textbf{0.1076}\\
\hline
GATv2 & 0.3252 & 0.1026 \\
GATv2 \& MP$_{i=1,2,3}$ & \textbf{0.3258} & \textbf{0.1220}\\
\hline
\end{tabular}
\label{table_multiview}
\end{center}
\end{table}

\subsection{Analysis of $\mathcal{L}_j$, $\mathcal{L}_s$, $\mathcal{L}_y$ (\bf Q3)}

In order to analyze the contribution of representation raw portfolio agreement, multi-view consistency and course-platform alignment, we define six variants of our proposed model: 
(1) MI-IaGRL-J, which adds $\mathcal{L}_j$ to the base model;
(2) MI-IaGRL-S, which adds $\mathcal{L}_s$ to the base model;
(3) MI-IaGRL-Y, which adds $\mathcal{L}_y$ to the base model;
(4) MI-IaGRL-J,S, which adds $\mathcal{L}_j$ and $\mathcal{L}_s$ to the base model;
(5) MI-IaGRL-J,Y, which adds $\mathcal{L}_j$ and $\mathcal{L}_y$ to the base model;
and (6) MI-IaGRL-S,Y, which adds $\mathcal{L}_s$ and $\mathcal{L}_y$ to the base model.

\begin{figure}[htbp]
\centering
\subfigure[Accuracy.]{
\begin{minipage}[t]{0.498\linewidth}
\centering
\includegraphics[width=1\linewidth]{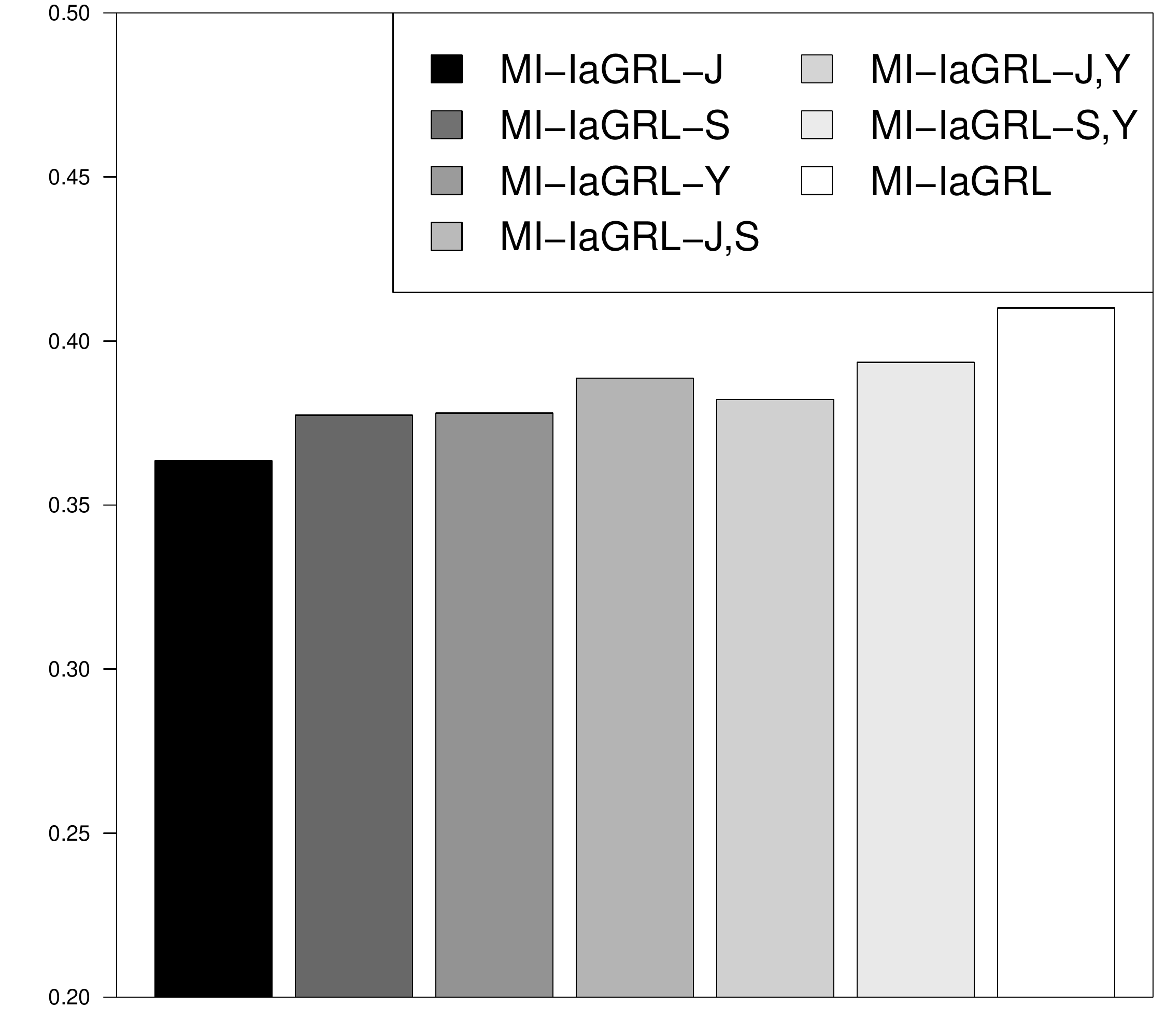}
\label{acc}
\end{minipage}%
}%
\subfigure[Macro-F1.]{
\begin{minipage}[t]{0.498\linewidth}
\centering
\includegraphics[width=1\linewidth]{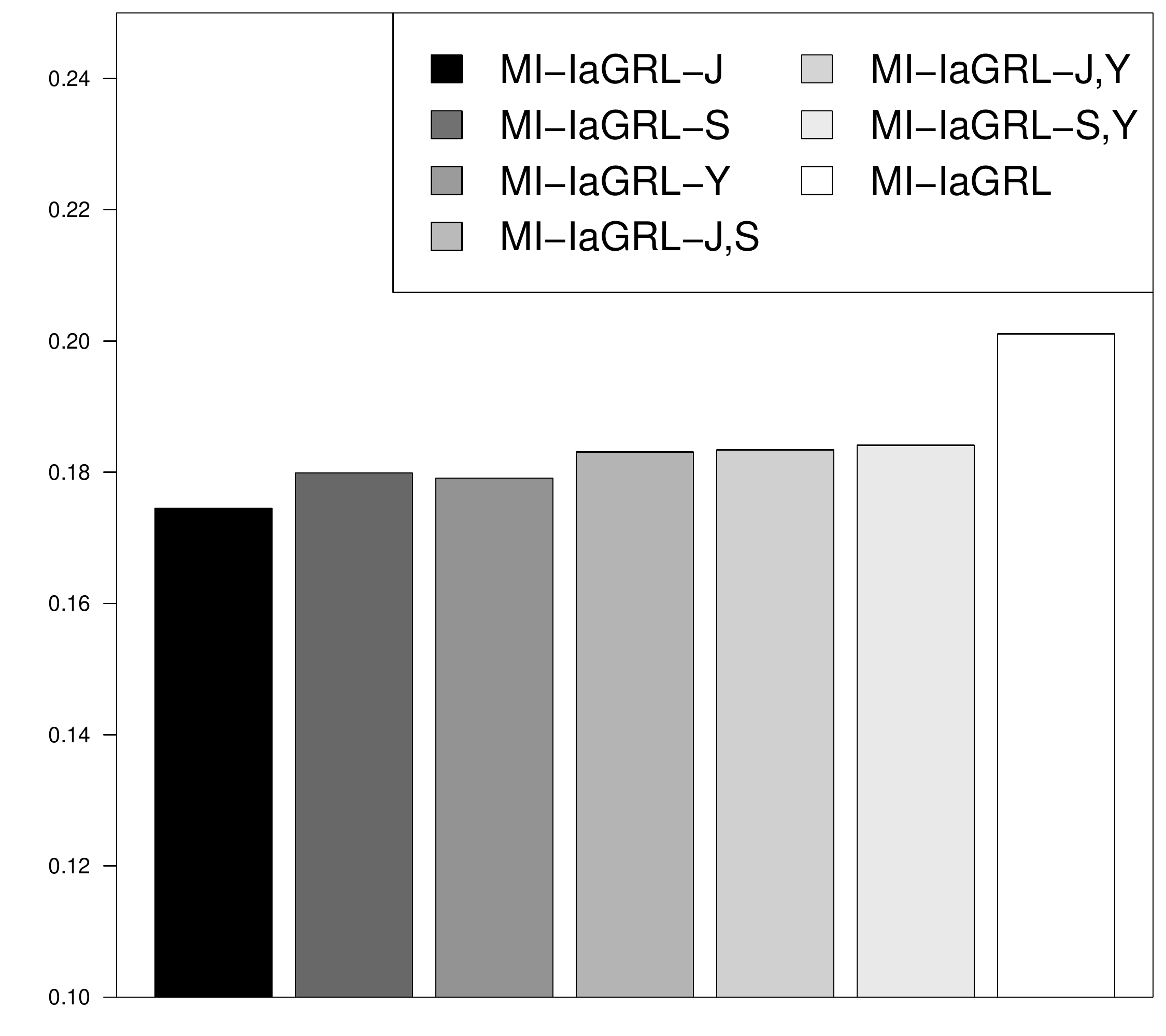}
\label{f1}
\end{minipage}%
}%

\caption{Analysis of $\mathcal{L}_j$, $\mathcal{L}_s$, $\mathcal{L}_y$.}
\label{figure_mi}
\end{figure}

As shown in Figure~\ref{figure_mi}, we compare the MI-IaGRL-J, MI-IaGRL-S, MI-IaGRL-Y, MI-IaGRL-J,S, MI-IaGRL-J,Y, MI-IaGRL-S,Y and MI-IaGRL in the experiment.
When the combination of loss functions increases from single to two to three, the overall trend of metric values are increasing.
The results indicate that the integrated raw portfolio-representation agreement, the multi-view consistency, and    the course-platform alignment significantly improve the performance of MOOC quality evaluation.

\section{Related Work}

Our work is related to the following two domains of prior work, including MOOC quality evaluation and graph representation learning.

\subsection{MOOC Quality Evaluation}

Our work has a connection with MOOC quality evaluation. 
Prior literature on MOOC quality evaluation lies in two aspects:
(1) manual evaluation~\cite{DBLP:journals/iahe/WangLLMY21}, and (2) automated evaluation~\cite{DBLP:journals/uais/Perez-MartinRM21,DBLP:journals/aai/BetanzosCB17}.
Manual-based methods evaluate the course quality by domain experts based on a pre-defined rubric.
For example, Wang {\it et al.} discussed the quality analysis of instructional design based on the ten-principle framework~\cite{DBLP:journals/iahe/WangLLMY21}.
The manual-based method can evaluate course quality accurately, but they are not suitable for large-scale applications.
More and more automatic methods appear, such as, 
Zhuo {\it et al.} designed a teaching quality assessment model on the MOOC platform based on comprehensive fuzzy evaluation~\cite{DBLP:journals/ijet/ZhuoD17}.

\subsection{Graph Representation Learning}

Different from the traditional graph optimization method~\cite{chen2022improved,wang2020sccwalk} focusing on efficiency, the graph embedding method focuses on information extraction.
Graph embedding aims to project nodes in a graph into a $d$-dimensional vector space, in which the representation of nodes can reflect the relationship between nodes, and retain the semantic information of nodes. 
Graph embedding methods can be categorized into homogeneous graph embedding(node2vec~\cite{DBLP:conf/kdd/GroverL16}, struct2vec~\cite{DBLP:conf/kdd/RibeiroSF17} and Deepwalk~\cite{DBLP:conf/kdd/PerozziAS14}), heterogeneous graph embedding(metapath2vec~\cite{DBLP:conf/kdd/DongCS17}, HHNE~\cite{DBLP:conf/aaai/WangZS19a}, SHNE~\cite{DBLP:conf/wsdm/ZhangSC19}).
However, the above methods ignore the mutual information.
In order to handle the information-aware of graphs, there are several methods have been proposed, including DGI~\cite{DBLP:conf/iclr/VelickovicFHLBH19}, HDGI~\cite{DBLP:journals/corr/abs-1911-08538}.

\section{Conclusion}

In this paper, we study the problem of MOOC course quality evaluation with MOOC heterogeneous information networks and propose an information-aware graph representation learning framework for multi-view MOOC quality evaluation.
Specifically, we first formulate the problem of MOOC quality evaluation as a multi-view graph representation learning task.
Second, we construct MOOC HIN and propose to exploit meta-paths to extract the semantics of MOOC relationships from different views.
Third, we identify three types of validity of course representations,  with (i) the agreement on expressiveness between the raw course portfolio and the learned course representations; (ii) the consistency between the representations in each view and the unified representations; and (iii) the alignment between the course and MOOC platform representations.
Finally, we conduct extensive experiments over real-world MOOC data to validate the effectiveness of our method.

\section{Acknowledgments}
This work is supported by the Fundamental Research Funds for the Central Universities 2412019ZD013, NSFC (under Grant No.61976050, 61972384, 61806050, 62106040), Jilin Science and Technology Department 20200201280JC, 
the Science and Technology Development Fund, Macau SAR (File no. SKL-IOTSC-2021-2023 to Pengyang Wang), the Start-up Research Grant of
University of Macau (File no. SRG2021-00017-IOTSC to Pengyang Wang).

\bibliography{aaai23}

\end{document}